\documentclass[prd
,floatfix,nofootinbib]{revtex4}
\usepackage{bm}
\usepackage{amssymb}
\usepackage{graphicx}
\usepackage{epstopdf}

\begin{document}

\title{Transversality of Electromagnetic Waves in the Calculus--Based Introductory Physics Course}

\author{Lior M.~Burko}

\affiliation{Department of Physics and Center for Space Physics and Aeronomic Research, University of Alabama in Huntsville, Huntsville, Alabama 38599, USA}

\date{August 7, 2008}

\begin{abstract}
Introductory calculus--based physics textbooks state that electromagnetic waves are transverse and list many of their properties, but most such textbooks do not bring forth arguments why this is so. Both physical and theoretical arguments are at a level appropriate for students of courses based on such books, and could be readily used by instructors of such courses. Here, we discuss two physical arguments (based on polarization experiments and on lack of monopole electromagnetic radiation), and the full argument for the transversality of (plane) electromagnetic waves based on the integral Maxwell equations. We also show, at a level appropriate for the introductory course, why the electric and magnetic fields in a wave are in phase and the relation of their magnitudes. 
\end{abstract}


\maketitle

\section{Introduction}

Electromagnetic waves in free space constitute an important part of the calculus--based introductory physics course. In addition to other properties, all the textbooks we have surveyed include a discussion of the transverse nature of the waves. Specifically, that  (plane) electromagnetic waves have the electric field strength ${\bf E}$ and the magnetic induction ${\bf B}$ perpendicular to the propagation direction of the wave and to each other so that $({\bf E}/E,{\bf B}/B,{\bf S}/S)$ is an even permutation of $({\hat x},{\hat y},{\hat z})$, where ${\bf S}$ is the Poynting vector. 

The extent of the coverage and the arguments brought in the textbooks vary widely. While we have not checked all textbooks, none of the ones we have surveyed include a full and detailed theoretical argument for these important properties, even though all textbooks include the infrastructure needed for that purpose, i.e., the (integral) Maxwell equations. Most textbooks derive the wave equations for the transverse components of the ${\bf E}$ and ${\bf B}$ fields, but do not show that these wave equations are unique, i.e., that a longitudinal component violates the Maxwell equations. That is, they show that transverse waves are {\em consistent} with the Maxwell equations, but not that they are required. We make the case here that the theoretical argument is important for students of the calculus--based introductory course. Much research has shown that understanding and retention of physics is enhanced if important concepts are not brought just as factoids, but rather derived from theoretical and physical argumentation. In the case of transversality of (plane) electromagnetic waves our case is even strengthened, as the argument is certainly at the right level for students of such courses. Indeed, a general demonstration without assuming planar symmetry is clearly beyond the scope of the introductory course, and rightfully belongs to more advanced courses. However, the physics and mathematics needed for the demonstration for plane electromagnetic waves are readily available for students of the introductory course, and in fact are already included in most textbooks we surveyed. Therefore, no new physics or mathematics are needed to demonstrate the transversality of (plane) electromagnetic waves in the introductory course. All that is needed is to arrange the arguments that are already in the student's toolkit to achieve deeper level of understanding. 

The textbooks we have surveyed (and again, we emphasize our literature search was not exhaustive, as the number of textbooks is indeed very large; we do believe we have found the tendencies of existing textbooks) fall under the following categories. Category A: textbooks that postulate the transversality (but don't derive it), and then show consistency with the (either all four or just the dynamical) Maxwell equations \cite{cat_A}; category B: textbooks that only use the two dynamical equations (Faraday's law and the Amp\`{e}re--Maxwell law) to derive the wave equations for the (transverse) components of the ${\bf E}$ and ${\bf B}$ fields, but don't show that {\em only} transverse fields are allowed \cite{cat_B1,cat_B2}; and category C: textbooks that add to the treatment of category B also a demonstration that a longitudinal component of the ${\bf E}$ field is disallowed (but not that a transverse component is consistent with the Maxwell equations) \cite{cat_C}. Interestingly, even introductory books that discuss the differential Maxwell equations do not include a demonstration of the transversality of electromagnetic waves  \cite{cat_D,Feynman} (we have not surveyed more advanced books, appropriate for upper division courses on electromagnetism). 

None of the textbooks we have surveyed include physical arguments for the transversality of electromagnetic waves, even though all of them discuss polarization. On the other hand, none of the textbooks we have surveyed include a discussion of the impossibility of genuine spherical electromagnetic waves (either monopole of a combination of multipoles) \cite{comay}. Either polarization or hypothetical spherical electromagnetic waves may be used in physical arguments for the transversality of electromagnetic waves. This omission is surprising, as building strong physical intuition is widely considered a primary goal of such courses. 

Many of the properties of electromagnetic waves can be demonstrated also for the algebra--based introductory course \cite{carr}, but examination of these arguments and their use in textbooks are beyond the scope of this Paper. 

Our theoretical argument is by no means original, and surely has been made numerous times. It is however absent in its entirety from all textbooks we have surveyed. Even if it does appear in textbooks we have not surveyed, we believe it is fair to state that most textbooks do not include the full argument, and that similarly most calculus--based introductory physics courses do not discuss it. 

In advanced courses the transversality of electromagnetic waves is easily shown \cite{jackson} from the  Maxwell equations and the wave equation for the vector potential $A_{\mu}$ in the Lorenz gauge, and the skew symmetry of the Maxwell--Faraday tensor $F_{\mu\nu}:=\,\partial_{\mu}A_{\nu}-\,\partial_{\nu}A_{\mu}$. Consider, say, a vector potential  
$A_{\mu}=A\,\cos(kz-\omega t)\,\delta^x_{\mu}$. The only nonvanishing independent components of $F_{\mu\nu}$ are $F_{tx}=E_x$ and $F_{xz}=B_y$, and $E_x=(\omega /k)B_y$.  These ${\bf E}$ and ${\bf B}$ fields manifestly satisfy the aforementioned properties, although the formality of the demonstration ---in addition to being inappropriate for the introductory course--- is devoid of physical insight. (One may of course try to impose longitudinality: take $A_{\mu}=A\,\cos(kz-\omega t)\,\delta^z_{\mu}$. One then finds that $F_{tz}$ is non-zero, and one might be tempted to identify it with $E_z$. However, the postulated vector potential is not a solution for the Maxwell equations in the Lorenz gauge, as can be readily verified.)  One can also show the transversality without invoking tensor analysis, using only vector analysis methods \cite{landau}. 
The demonstration using elementary methods is important, as it involves more physical insight than the advanced, more formal proof. 

Notably, electromagnetic waves in matter do not have to be transverse \cite{halevi}. Longitudinal electromagnetic waves (or combinations of longitudinal and transverse waves) indeed exist in inhomogeneous media and in cavities (because of the cavity's boundary conditions) \cite{jackson}. One may also have in vacuum electromagnetic waves with ${\bf E}\times{\bf B}=0$, but those are standing waves (with vanishing Poynting vector) \cite{shimoda}. These interesting cases are typically beyond the scope of the introductory course.

In this Paper we present a full argument for the transversality of (plane) electromagnetic waves in vacuum. We are hopeful that instructors of calculus--based introductory physics courses present a full argument in their classes. We use SI units despite their awkwardness (for the description of electrodynamics), as this appears to be the nearly universal practice in textbooks. We use only the integral form of the Maxwell equations, as nearly all introductory textbooks (for {\em calculus}--based physics) refrain from introducing differential operators.  In Section \ref{phys_arg} we describe physical arguments for the transversality of electromagnetic waves, based on polarization and on (lack of ) monopole radiation. In Section \ref{theor_arg} we present a full argument at a level appropriate for the introductory calculus--based physics course, as follows: First, we show that ${\bf E}$  cannot have a longitudinal component, and that ${\bf E}$ may have a transversal component. We then repeat the arguments for the ${\bf B}$ field. We then show that the ${\bf E}$ and ${\bf B}$ fields are orthogonal to each other and to the direction of propagation, and finally, we show that the ${\bf E}$ and ${\bf B}$ fields satisfy the wave equation. We emphasize that the only assumption we make is that of planar symmetry. Except for this assumption (which makes the derivation possible for the introductory course) our argument is general.

\section{Physical arguments for transversality}\label{phys_arg}

\subsection{Polarization}

Many introductory courses include a demonstration of polarization that includes two polaroid films, so that one can be rotated relative to the other. When the polarization directions of the two polaroids are orthogonal, no light is going through the polaroids, and the image on the screen is dark. 

Assume there were a longitudinal component to the ${\bf E}$ field. This component may be transmitted through either polaroid, so that a longitudinal component would arrive to the screen through both polaroids even when they are orthogonal. Therefore, with a longitudinal component the screen will never get totally dark when the two polaroids are in a plane perpendicular to the direction of propagation of the light. A variant of this argument was given by Schutz \cite{schutz}. Schutz considers the two polarizers rotating relative to each other in the plane orthogonal to the direction of propagation. As the brightness on the screen oscillates with the polarizers, one must conclude the electric field acts across its motion, i.e., transversally. 

In order to make our argument compelling, students need to have at least a qualitative understanding of how simple dichroic devices such as a wire--grid polaroid work (most introductory texts do not discuss in depth other polarizers, such as 
beam--splitting polarizers, birefringent polarizers, or polarization by reflection). Such understanding may be expected from students of the relevant course: there are three microscopic mechanism that filter out the component of the ${\bf E}$ field in the plane of the polaroid parallel to the direction of the wires (or polyvinyl alcohol polymers doped with an iodine solution in a commercial polaroid): first, by the Lorentz force law, conduction electrons are accelerated along the wires, and oscillate with the oscillating ${\bf E}$ field. When colliding with lattice atoms, kinetic energy is transferred to the lattice, thereby transferring energy from the parallel component of the ${\bf E}$ field to the lattice (``Joule heating"). Second, the oscillating conduction electrons reradiate in all directions, so that half the reradiation is backward (reflection) and the remainder forward. The backward reradiation takes away more energy from the parallel component of the ${\bf E}$ field. The forward reradiation is not entirely in the direction of the original incident plane wave, so that it scatters and little of it arrives at the detector. Most importantly, the forward reradiation is generally  out of phase with the incident ${\bf E}$ field (phase difference of $\pi$ radians), thereby reducing it further by destructive interference \cite{pedrotti}. 

Some textbooks explain the polarization effect by stretching a mechanical model too far. In such texts the electric field is modeled by a tension wave in a string that is passing through a fence with parallel beams \cite{cat_B2}. The mechanical wave naturally is transmitted only if the string oscillates parallel to the beams. This argument might imply the wrong component of the ${\bf E}$ field is transmitted. For this reason, those texts refer to the polarization direction of the polarizer, stating that perpendicular components are filtered out. But this argument may then persuade some students to believe a longitudinal component of the ${\bf E}$ field is also filtered out, such that this sort of argumentation, in addition to not empowering the students with an understanding of simple dichroic devices work, also might prevent them from understanding a simple physical argument for why electromagnetic waves are transverse, and give the students a misleading physical picture of electromagnetic fields oscillating in the space between the wires.

\subsection{No monopole radiation}

Consider a hypothetical monopole electromagnetic wave. Then, a radially pulsating spherical charge distribution would emit such spherical outgoing waves. As the fundamental equations of physics are symmetrical under time reversal ($t\to -t$), an incoming imploding spherical monopole wave would make an otherwise static spherical charge distribution pulsate radially. Each charge pulsates radially because of a radial force acting on it. 
The form of the Lorentz force law then implies there were a radial ${\bf E}$ field acting on the charges. But a radial ${\bf E}$ field is perpendicular to  the imploding spherical wave front and therefore in the same direction of its motion, so that the ${\bf E}$ field would have a longitudinal component. This qualitative argument implies that a monopole electromagnetic wave is necessarily longitudinal. (While students of the introductory course generally have not studied multipole expansions of the electromagnetic field, the argument may still be used in reference to a truly spherical wave front.) 

We do know, however, that outside a spherical charge distribution the electromagnetic field is static, as the monopole piece of the electromagnetic field is non-radiative. Students are exposed to this argument also in Newtonian gravity: because of the inverse square force law (common to both Newton's and Coulomb's laws) the field strength of any spherical charge (or mass) distribution is the same as if all the charge (or mass) were concentrated at the center. Specifically, the electric field outside a radially pulsating spherical charge distribution is static. 
Therefore, the longitudinal component of the ${\bf E}$ field cannot exist, as such a longitudinal component would have to be radiated by a pulsating spherical charge distribution \cite{schutz}. 

This physical argument may be cast in a more mathematical form, using a theorem from vector calculus that states that no unit  
two--dimensional continuous vector field ${\bf V}$ on a sphere may exist on the entire sphere ($0\le\theta\le\pi$, $0\le\phi <2\pi$)  \cite{rosenthal}. This theorem is normally beyond the scope of the introductory course. In addition, it requires the assumption of transversality (namely, that ${\bf V}$ is a two-dimensional vector field on the sphere, or ${\bf V}\cdot{\bf r}=0$). However, it may be used to argue that any such spherical electromagnetic wave would have to be longitudinal.

\section{Construction of the transversality argument, and deriving the properties of the waves}\label{theor_arg}

\subsubsection{Show that ${\bf E}$ does not have a longitudinal component}
\begin{figure}
 \includegraphics[width=3.4in]{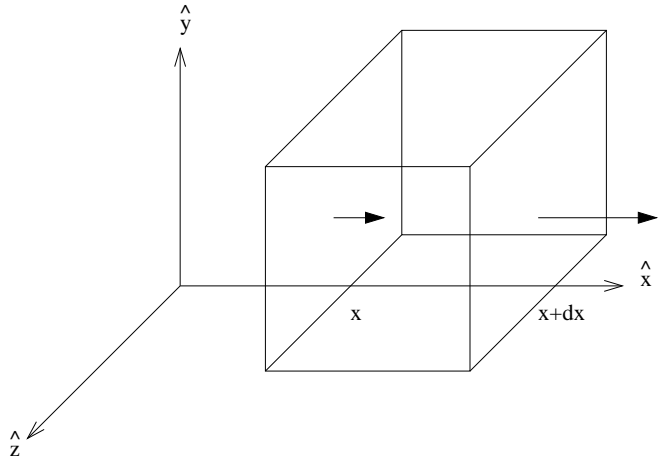} 
\caption{A gaussian surface constructed in a shape of a rectangular box whose faces parallel the axes. The ${\hat x}$ components of either ${\bf E}$ or ${\bf B}$ have to be independent of $x$. The arrows indicate a case in which they are not. Then, the surface integral must be nonzero, and proportional to their $x$ derivatives. In the case of ${\bf E}$ this implies free charges inside the box, i.e., a continuous charge distribution, which contradicts the vacuum assumption. In the case of ${\bf B}$ that distribution would be of magnetic monopoles.}
\label{gauss1}
\end{figure}

Consider a plane electromagnetic wave traveling in vacuum in the direction of ${\hat x}$. (Our coordinates may be rotated to this orientation.) Because of the planar symmetry the most general electric field strength is ${\bf E}={\bf E}(t,x)$. In what follows we omit the explicit time dependence of fields. We first show there can be no component $E_x$: construct a gaussian surface enclosing a volume $V$ as in Fig. \ref{gauss1}. According to Gauss's law, 
$$\oint_{\,\partial V}{\bf E}\cdot d{\bf A}=\frac{q}{\epsilon_0}\, ,$$
where $q$ is the total charge inside the gaussian surface, and $\epsilon_0$ is the permittivity of vacuum. Here, $\,d{\bf A}$ is a  surface element normal to the (orientable) surface $\,\partial V$, defined conventionally such that it is positive when pointing outward. 
Calculate the integral on the LHS: take ${\bf E}=E_x(x){\hat x}+E_y(x){\hat y}+E_z(x){\hat z}$. Then 
$$\oint_{\,\partial V}{\bf E}\cdot d{\bf A}=[E_x(x+\,dx)-E_x(x)]\,dy\,dz=0$$
as there are no charges. The components $E_y$ and $E_z$ identically do not contribute to the LHS integral, as these components are not functions of $y$ or $z$, respectively, and as they are in the plane of the faces of the surface whose normals are in the direction of ${\hat x}$. Notice that there is no flux of $E_x$ through the faces of the gaussian surfaces in the ${\hat y}$ or ${\hat z}$ directions because of the orthogonality of $E_x$ to the surface's normals.  
This immediately implies 
$\,\partial E_x/\,\partial x=0$, or $E_x$ does not obey a non-trivial wave equation. Therefore, the most general electric field is ${\bf E}=E_y(x){\hat y}+E_z(x){\hat z}$. Without loss of generality, we may rotate our coordinate system so that ${\bf E}=E_y(x){\hat y}$.

Our demonstration depends on the lack of charges. Presence of charges requires a non-vanishing gradient of $E_x$. 
Indeed, Gauss's law would not be violated for longitudinal waves without assuming vacuum, as there are charges and inhomogeneities in matter. Therefore, this demonstration does not rule out longitudinal waves in matter or inhomogeneous media. 

\subsubsection{Show that ${\bf E}$ may have a transversal component}
\begin{figure}
 \includegraphics[width=3.4in]{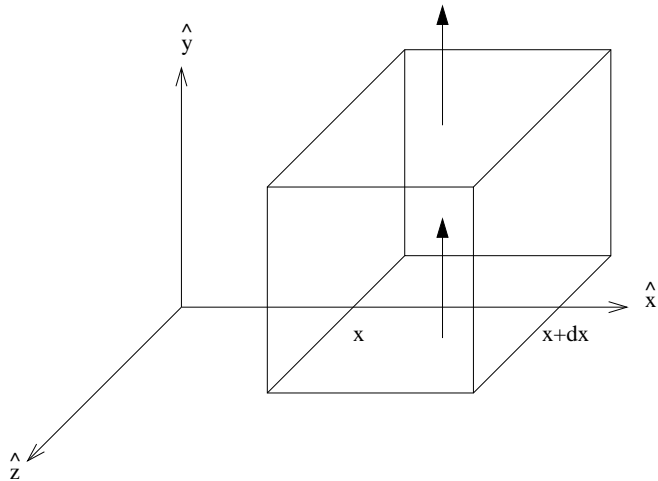} 
\caption{A transverse component to either ${\bf E}$ or ${\bf B}$ leads to no contradiction: as the fields are functions of only the spatial coordinate $x$ (in addition to the time $t$), the surface integral of the fields vanish along all faces, either because of the field being parallel to the face (faces parallel to the $x-y$ and $y-z$ planes) or because the field exiting the surface completely balances the field entering (faces parallel to the $x-z$ plane). It is therefore important that the fields are only functions of $x$, otherwise the cancellation would not be exact. }
\label{gauss2}
\end{figure}

Consider next an electric field ${\bf E}=E_y(x){\hat y}$, as in Fig.~\ref{gauss2}. Applying Gauss's law does not yield a restriction on $E_y$, because the LHS of the integral vanishes identically, as $E_y=E_y(x)$:
$$\oint_{\,\partial V}{\bf E}\cdot d{\bf A}=[E_y(x, y+\,dy)-E_x(x,y)]\,dx\,dz=0\, .$$

This step may of course be combined with the preceding one by taking ${\bf E}=E_x(x){\hat x}+E_y(x){\hat y}+E_z(x){\hat z}$, and showing that Gauss's law becomes $\,\partial E_x/\,\partial x=0$. In conclusion, a transverse component for ${\bf E}$ is consistent with Gauss's law. Of course, were there not a transverse component, the solution would become trivial (no electromagnetic wave). Therefore, for any electromagnetic wave a transverse ${\bf E}$ field is necessary.   

\subsubsection{Show that ${\bf B}$ does not have a longitudinal component}
We next use similarly the magnetic Gauss law, 
$$\oint_{\,\partial V}{\bf B}\cdot d{\bf A}=0\, .$$
The same argument as in Step 1 implies that $\,\partial B_x/\,\partial x=0$, so that also ${\bf B}$ does not have a longitudinal component. Therefore, ${\bf B}=B_y(x){\hat y}+B_z(x){\hat z}$. But we have no more freedom to rotate the coordinate system, a freedom which we have already exhausted in rotating it so that $E_z=0$. We may therefore not set any of the components of ${\bf B}$ equal to zero. As there are no magnetic monopoles, the ${\bf B}$ field has to be transverse also in matter.

\subsubsection{Show that ${\bf B}$ may have a transversal component}
Consider next a field ${\bf B}=B_y(x){\hat y}+B_z(x){\hat z}$. Applying Gauss's law does not yield a restriction on $B_y$ or on $B_z$, because the LHS of the integral vanishes identically, as $B_y=B_y(x)$ and $B_z=B_z(x)$:
\begin{eqnarray*}
\oint_{\,\partial V}{\bf B}\cdot d{\bf A}&=&[B_y(x, y+\,dy)-B_x(x,y)]\,dx\,dz\\
&+& [B_z(x, z+\,dz)-B_z(x,z)]\,dx\,dy=0\, .
\end{eqnarray*}

\subsubsection{Show that the ${\bf E}$ and ${\bf B}$ fields are orthogonal to each other and to the propagation direction}
\begin{figure}
 \includegraphics[width=3.4in]{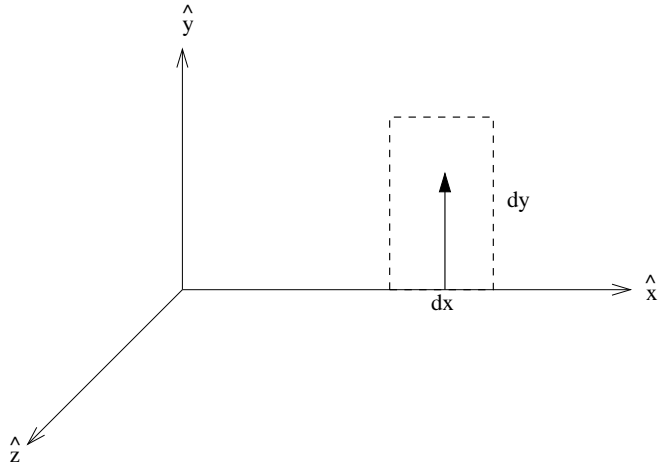} 
\caption{An Amp\`{e}re loop constructed so that the ${\hat z}$ component of the ${\bf B}$ field is perpendicular to the entire loop, so that it does not contribute to the line integral around it. When applied to the Amp\`{e}re--Maxwell law, an electric field in the direction of ${\hat y}$ has zero flux through the loop. The line integral around the loop is proportional to the $x$ derivative of the ${\hat y}$ component of ${\bf B}$. When applied to the Faraday law, the $x$ derivative of the ${\hat y}$ component of ${\bf E}$ is proportional to the time derivative of the ${\hat z}$ component of ${\bf B}$. The arrow indicates a ${\hat y}$ component to either ${\bf E}$ or ${\bf B}$.}
\label{Ampere1}
\end{figure}

Consider now an Amp\`{e}re loop enclosing an area $A$ as in Fig.~\ref{Ampere1}. The Amp\`{e}re--Maxwell law is 
$$\oint_{\,\partial A} {\bf B}\cdot \,d{\bf S}=\mu_0\, i+\epsilon_0\mu_0\,\frac{\,d\Phi_E}{\,dt}\, ,$$ 
where $\mu_0$ is the permeability of free space, $i$ is the total conduction current through the loop, and $\Phi_{E}$ is the flux of the ${\bf E}$ field through it. Here, $\,d{\bf S}$ is an element of the curve $\,\partial A$ along the loop, conventionally taken to be positive counterclockwise. 
Doing the integral on the LHS, 
$$\oint_{\,\partial A} {\bf B}\cdot\,d{\bf S}=[B_y(x+\,dx)-B_y(x)]\,dy=\frac{\,\partial B_y}{\,\partial x}\,dx\,dy=0\, ,$$
as the $B_z$ component is perpendicular to the entire loop, and as there are no electric field flux through the loop (because the ${\bf E}$ field in in the direction of ${\hat y}$) and also no free currents. Therefore, 
$\,\partial B_y/\,\partial x=0$. Therefore, we find that ${\bf B}=B_z(x){\hat z}$. We therefore find that ${\bf E}\cdot{\bf B}=0$, i.e., the fields are orthogonal to each other and also to the direction of propagation ${\hat x}$. In addition, ${\bf E}\times{\bf B}=E_y(x)B_z(x)\,{\hat y}\times{\hat z}=E_y(x)B_z(x)\,{\hat x}$, i.e., the cross product of the fields is in the direction of propagation of the wave.

\subsubsection{Show that the ${\bf E}$ and ${\bf B}$ fields satisfy the wave equation with speed $c$}
This part of the argument appears in most textbooks we have surveyed \cite{cat_B1,cat_B2,cat_C}. However, without the preceding parts of the argument it is merely a demonstration of consistency of the transverse wave equation with the Maxwell equations, not their unavoidability. 

\begin{figure}[htbp]
 \includegraphics[width=3.4in]{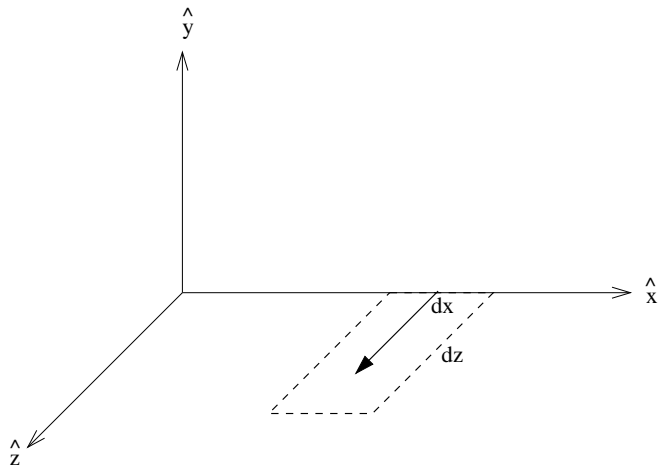} 
\caption{An Amp\`{e}re loop constructed so that the ${\hat z}$ component of the ${\bf B}$ field is in the plane of the loop. The line integral around the loop is proportional to the $x$ derivative of the ${\hat z}$ component of ${\bf B}$, which by Faraday's law is proportional to the time derivative f the electric field's flux through it. The arrow indicates the ${\hat z}$ component of the ${\bf B}$ field. The ${\bf E}$ field (which is in the direction of ${\hat y}$) is not shown.}
\label{Ampere2}
\end{figure}

We now use the loop in Fig.~\ref{Ampere2} to evaluate the Amp\`{e}re--Maxwell law. On the LHS, 
$$\oint_{\,\partial A} {\bf B}\cdot \,d{\bf S}=[B_z(x+\,dx)-B_z(x)]\,dz=\frac{\,\partial B_z}{\,\partial x}\,dx\,dz\, .$$
The flux of the ${\bf E}$ field is simply $\Phi_E=E_y(x)\,dx\,dz$, so that the RHS is evaluated to equal 
$$\epsilon_0\mu_0\frac{\,d\Phi_E}{\,dt}=\epsilon_0\mu_0 \frac{\,\partial E_y}{\,\partial t}\,dx\,dz\, ,$$ so that the Amp\`{e}re--Maxwell law yields 
\begin{equation}\label{eq1}
\frac{\,\partial B_z}{\,\partial x}=\epsilon_0\mu_0\frac{\,\partial E_y}{\,\partial t}\, .
\end{equation}

We next consider again the Amp\`{e}re loop in Fig.~\ref{Ampere1}, but this time apply it to the Faraday law, 
$$\oint_{\,\partial A} {\bf E}\cdot\,d{\bf S}=-\frac{\,d\Phi_B}{\,dt}\, .$$ Here, $\Phi_B$ is the flux of the ${\bf B}$ field 
through the loop. The LHS is evaluated as 
$$\oint_{\,\partial A} {\bf E}\cdot\,d{\bf S}=[E_y(x+\,dx)-E_y(x)]\,dy=\frac{\,\partial E_y}{\,\partial x}\, dx\,dy\, .$$
The RHS is $$-\frac{\,d\Phi_B}{\,dt}=-\frac{\,\partial B_z}{\,\partial t}\,dx\, dy\, ,$$
so that 
\begin{equation}\label{eq2}
\frac{\,\partial E_y}{\,\partial x}=-\frac{\,\partial B_z}{\,\partial t}\, .
\end{equation}

Many introductory texts \cite{cat_B1,cat_B2,cat_C} show that by differentiating Eqs.~(\ref{eq1}) and (\ref{eq2}) and eliminating mixed second derivative terms, one can obtain wave equations. With the identification $c^2=1/(\epsilon_0\mu_0)$ these equations lead to 
$$\frac{\,\partial^2 E_y}{\,\partial t^2}-c^2\,\frac{\,\partial^2 E_y}{\,\partial x^2}=0$$
and
$$\frac{\,\partial^2 B_z}{\,\partial t^2}-c^2\,\frac{\,\partial^2 B_z}{\,\partial x^2}=0\, .$$

\subsubsection{The magnitudes of the ${\bf E}$, ${\bf B}$ fields}

To show the relation between the magnitudes of the ${\bf E}$ and ${\bf B}$ fields we present an adaptation of an argument originally made by Feynman \cite{Feynman}. Consider an infinite planar current sheet, carrying current density ${\bf j}=-j\,{\hat y}$. A current sheet is a two dimensional surface carrying current, and can be approximated by a large number of wires, all carrying current in the same direction, so that the total current per unit length (perpendicularly to the wires) is $j=i/\ell$, where $i$ is the total current included in the length $\ell$. Current sheets are a very useful model in magnetohydrodynamics (MHD) and in heliophysics. 

\begin{figure}[htbp]
 \includegraphics[width=3.4in]{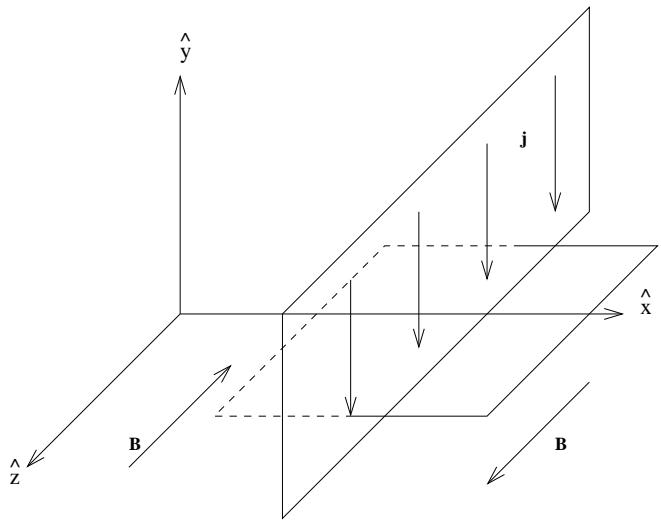} 
\caption{A current sheet in the $y-z$ plane, carrying current density ${\bf j}$. The direction of the ${\bf B}$ field is as indicated. The Amp\`{e}re loop is in the $x-z$ plane. }
\label{current_sheet}
\end{figure}

We first recall the ${\bf B}$ field of a stationary current sheet \cite{cat_D}. Consider a current sheet aligned as in Fig.~\ref{current_sheet}. To find the ${\bf B}$ field outside the current sheet we construct an Amp\`{e}re loop whose plane is perpendicular to the plane of the current sheet. The ${\bf B}$ field must be parallel to the plane of the current sheet, and perpendicular to the direction of the current. (One may be convinced of that result by considering the elementary problem of  current wire, and considering the current sheet as the limiting case of infinitely many such current wires.) As this problem is stationary, we may use the original form of the Amp\`{e}re law (no time--changing flux of an electric field), i.e., 
$$\oint_{\,\partial A} {\bf B}\cdot \,d{\bf S}=\mu_0\, i\, ,$$ 
where $i$ is the total current going through the loop, whose length in the $z$--direction is $\,\Delta z$. Because the current sheet is infinite, the magnitude of the field strength $|{\bf B}|:=B$ is constant along the legs of the loop in the $z$--direction. 
The LHS of the Amp\`{e}re law then is $\oint_{\,\partial A} {\bf B}\cdot \,d{\bf S}=-2B\,\Delta z$, and the RHS is simply 
$\mu_0\, i=-\mu_0\,j\,\Delta z$. Putting the two expressions together, one finds that $B=\frac{1}{2}\mu_0\,j$. Notably, $B$ is independent of the distance from the current sheet.

\begin{figure}[htbp]
 \includegraphics[width=3.4in]{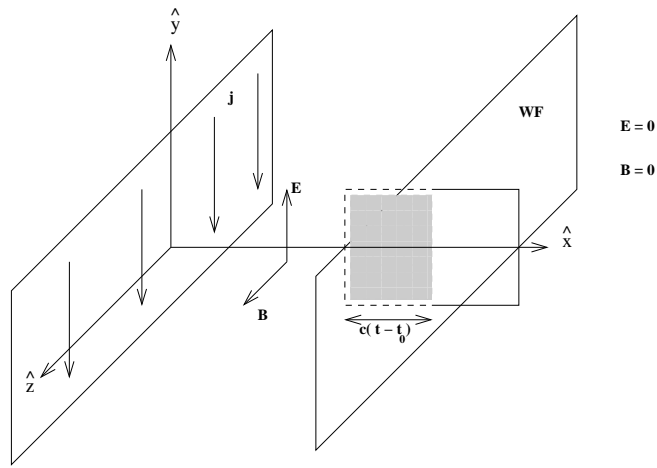} 
\caption{A current sheet in the $y-z$ plane, carrying current density ${\bf j}$ as in Fig.~\ref{current_sheet}. The current sheet is turned on abruptly, so that plane waves propagate with speed $c$ to the right of the current sheet in the ${\hat x}$ direction, and to its left in the $-{\hat x}$ direction. The wavefront (WF) is shown only to the right of the current sheet. At time $t_0$ the wave front touched the dashed segment in the $-{\hat y}$ direction of the Amp\`{e}re loop. The space between the wave front and the current sheet has uniform ${\bf B}$ field, and ahead of the wavefront the ${\bf B}$ (and ${\bf E}$) field vanishes. The only part of the loop with non-vanishing ${\bf B}$ field is behind the wave front, shown shaded.}
\label{wave_front}
\end{figure}

Consider next a current sheet that is abruptly turned on at time $t=0$. That is, ${\bf j}=-j\,\Theta(t)\,{\hat y}$, where $\Theta(t)$ is the Heaviside step function. Before the current is turned on, the ${\bf B}$ field is zero everywhere. Following the abrupt turning on of the current, a non--zero ${\bf B}$ field is starting to fill up space, the wavefront propagating at the speed $c$. Therefore, we have a propagating wavefront, so that in front of the wavefront the ${\bf B}$ field vanishes, and behind it is uniform as shown above. To find the accompanying ${\bf E}$ field we use the Faraday law for the Amp\`{e}re loop in Fig.~\ref{wave_front}. Only the part of the Amp\`{e}re loop behind the wavefront has non-vanishing ${\bf B}$--field flux. As the wavefront is propagating  with speed $c$, the area of the loop where there is a ${\bf B}$ field is increasing, so that there is a time--changing ${\bf B}$--field flux through the loop. The ${\bf E}$ field between the wavefront and the current sheet is parallel to the current sheet, and counter-parallel to the current. That is, ${\bf E}=E\,{\hat y}$. (See Fig.~\ref{wave_front}.) 
Using the Faraday law, 
$$\oint_{\,\partial A} {\bf E}\cdot\,d{\bf S}=-\frac{\,d \Phi_B}{\,dt}\, ,$$
one finds that the RHS equals $\oint_{\,\partial A} {\bf E}\cdot\,d{\bf S}=-E\,\Delta y$, and the flux of the ${\bf B}$ field through the loop at the time $t> t_0$ is $\Phi_B=B\,c(t-t_0)\,\Delta y$, so that the RHS of the Faraday law equals 
$-\,d\Phi_B/\,dt=cB\,\Delta y$. Setting the two hand sides equal to zero, we find $E\,\Delta z=cB\,\Delta z$, or
$$E=cB\, .$$

Although our argument was made only for  the configuration of a current sheet that is abruptly turned on, the result that $E=cB$ for an electromagnetic wave is general. The general result, however, is beyond the scope of the introductory course. An important consequence is that in an electromagnetic wave the ${\bf E}$ and ${\bf B}$  fields are in phase. That is, as at any given time $E=cB$, when $E$ is maximal so is $B$, when $E$ vanishes $B$ is zero too, etc. The result that $E=cB$ appears in many introductory texts (e.g., in \cite{cat_B2}). However, in such books an extra assumption is typically made, namely that the fields are harmonic and are in phase.  Students of the introductory course, who normally are not familiar with Fourier theory, often find it unconvincing to base an important general physical result on the mathematical properties of sinusoidal waves, in addition to having to memorize yet another factoid, namely that the fields are in phase.

At this point one is in a position to make the following discussion. We found that $E=cB$. this means that the ratio $E/B$ for an electromagnetic wave equals $c$, which depends in magnitude on the system of units. E.g., in SI units this ratio is $3\times 10^8\,{\rm m}\,{\rm s}^{-1}$, and in cgs units is it $3\times 10^{10}\,{\rm cm}\,{\rm s}^{-1}$. It is therefore natural to use units that put the ${\bf E}$ and ${\bf B}$  fields on equal footing, that is use units in which distance is measured in light-seconds. In such units the speed of light is $c=1$ (light-)second/second, or simply $c=1$. That is, in these units $E=B$. This choice of units becomes further motivated physically when one considers the energy density ${\cal E}$ in the electromagnetic field (not necessarily that of an electromagnetic wave), ${\cal E}=\frac{1}{2}\epsilon_0 (E^2+c^2B^2)$. For an electromagnetic wave $E=cB$, so that the energy densities stored in the ${\bf E}$ and ${\bf B}$ fields are equal. It is therefore suggestive to make the two fields symmetrical by using units in which $c=1$. It is often instructive to remind students that SI or cgs units were developed because of their convenience in describing everyday phenomena involving humans, and while all unit systems are in principle equivalent, these are not necessarily the units in which physical phenomena take their simplest or most natural form.

We therefore showed that electromagnetic waves are transverse, and that the electric and magnetic fields are perpendicular to each other. Also the directions are so that ${\bf E}\times{\bf B}$ is in the direction of propagation. We have also shown they have the same speed $c$, that they are in phase, and that the magnitudes of the  ${\bf E}$ and ${\bf B}$ is related by $E=cB$. This demonstration makes use of only the integral Maxwell equations, and is appropriate for the level of most calculus--based physics courses, and includes only arguments already available for the perspective students. We believe its main strength is that it avoids giving the students factoids without deep understanding, and instead empowers the student to gain deeper insight.

\section*{Acknowledgments}

The author wishes to thank Richard Price for discussions. 
This work has been supported by NASA/GSFC grant No.~NCC5--580 and by NASA/SSC grant No.~NNX07AL52A, and by NSF grant No.~PHY--0757344.

\end{document}